\documentclass[12pt]{article}

\usepackage{graphics,epsf}

\newcommand{\be}{\begin{equation}}
\newcommand{\ee}{\end{equation}}
\newcommand{\ba}{\begin{eqnarray}}
\newcommand{\ea}{\end{eqnarray}}

\def\reff#1{(\ref{#1})}

\def\spose#1{\hbox to 0pt{#1\hss}}
\def\ltapprox{\mathrel{\spose{\lower 3pt\hbox{$\mathchar"218$}}
 \raise 2.0pt\hbox{$\mathchar"13C$}}}
\def\gtapprox{\mathrel{\spose{\lower 3pt\hbox{$\mathchar"218$}}
 \raise 2.0pt\hbox{$\mathchar"13E$}}}

\begin{document}

\title{Equation of State for Spin Systems \\
with Goldstone Bosons: the $3d$ $O(4)$ Case}

\author{Attilio Cucchieri and Tereza Mendes\\[1mm]
{\em \normalsize IFSC -- S\~ao Paulo University,
C.P. 369, 13560-970 S\~ao Carlos SP, Brazil}}

\maketitle

\begin{abstract}
We propose an improved parametric form for the equation
of state of three-dimensional $O(N)$ spin systems. The proposed
form is a series expansion with two sets of terms, which contribute 
(mainly) separately to the description of the high- and low-temperature 
regions of the phase diagram. Our goal is a better description
of the low-temperature phase at zero magnetic field (i.e.\ 
the coexistence line), characterized by singularities induced
by Goldstone modes.
We test our proposed form by comparison
with existing Monte Carlo data for the $N=4$ case, which is 
of interest in studies of the QCD phase transition 
and for which the Goldstone-mode effects are quite pronounced.
We find that the description of the numerical equation
of state is indeed improved with respect to other fitting forms. 
In all cases considered we determine the coefficients
nonperturbatively, from fits to the data.
As a consequence, we are able to obtain a very precise 
characterization of the pseudo-critical line for the model.
\end{abstract}

\section{Introduction}

The $O(N)$ (or, more specifically, the $N$-vector) spin models correspond 
to a generalization of the Ising model to the case of the continuous
symmetry of rotation. The spin variables ${\bf S}_i$ are taken as
vectors on a sphere of unit radius in an $N$-dimensional space.
We consider $N \geq 2$. The Hamiltonian is defined in terms of the 
scalar product of nearest-neighbor spins on a three-dimensional square 
lattice as
\be
\beta\,{\cal H}\;=\;-J \,\sum_{<i,j>} {\bf S}_i\cdot {\bf S}_j
         \;-\; {\bf H}\cdot\,\sum_{i} {\bf S}_i \;,
\ee
where $J>0$ represents the ferromagnetic coupling and ${\bf H}$
the external magnetic field.

These models are of general interest for the statistical mechanics
of phase transitions \cite{Zinn-Justin}.
The $N=2$ case (also known as the $XY$ model) describes the superfluid
transition in liquid helium and the $N=3$ case corresponds to the
classical version of the Heisenberg model for ferromagnets.\footnote{The
$N=0$ and $N=1$ (the Ising model) cases, not considered here, correspond
respectively to models for the statistical properties of long polymers and
for the liquid-vapor transition in several fluid systems.}
Moreover, it is believed that the $N=4$ case describes the chiral
phase transition in finite-temperature QCD with two degenerate light quark 
flavors, which makes this class of models interesting to high-energy 
physics as well.\footnote{Two-dimensional $O(N)$ models are also of 
indirect interest in quantum field theories, as toy models for 
asymptotically-free gauge theories.}
In this case, the magnetization and the magnetic field of the spin
model correspond respectively to the chiral condensate and to the
quark mass for the QCD analogue of the transition
\cite{Pisarski:ms,Rajagopal:1992qz,Berges:1997eu}.

The $O(N)$ symmetry is exact in the Hamiltonian for $H=0$,
just like the reflection symmetry for the Ising model.
The main difference with respect to the Ising case is the possibility
of configurations where the spins are locally aligned but for long
distances this alignment is lost, yielding a null average for the
magnetization. Such configurations --- called {\em spin waves} ---
possess arbitrarily low energy and tend to destroy the 
order of the system even at low temperatures.
In fact, as opposed to the Ising model, the $O(N)$ models do not
display a phase transition with spontaneous magnetization\footnote{For
the case $N=2$ there is a phase transition of the Kosterlitz-Thouless 
type, without spontaneous magnetization.} in $d=2$.
In $d=3$ a phase transition occurs, with the presence of
spontaneous magnetization below the critical temperature.
The breaking of the (continuous) rotational symmetry at 
low temperatures, signaled by the spontaneous magnetization, is 
associated with Goldstone modes, the spin waves. These modes
cause the divergence of the zero-field susceptibility not only
at the critical temperature, but for the entire low-temperature
phase \cite{Zinn-Justin,Parisi}.
Note that the magnetic field defines a 
privileged direction in spin space and the
magnetization $M$ is the expectation value 
of the spin component along ${\bf H}$. There are
thus $N-1$ massless Goldstone modes, corresponding to the $N-1$
transverse spin components.

Spin models in the $O(N)$ class have been extensively studied using
analytic and numerical methods (see \cite{Pelissetto:2000ek}
for a recent review).
In particular, the nonperturbative study by Monte Carlo simulations 
is very efficient for these models due to the Swendsen-Wang cluster
algorithms \cite{Swendsen:ce}, which can be applied to the continuous-spin 
case by means of the embedding technique introduced by Wolff 
\cite{Wolff:1988uh}.
This study is important to test the perturbative predictions and 
to investigate cases for which these predictions are not available, 
or cannot be done with great accuracy. These problems include properties 
of the models in the presence of magnetic field and the direct calculation 
of long-distance observables such as the correlation length.
For example, the predicted singular behavior of the longitudinal
susceptibility 
for vanishing $H$ ---
mentioned above and induced by Goldstone modes at low temperatures ---
was directly observed in Monte Carlo simulations of the cases $N =$ 2, 4, 6
respectively in Refs.\ \cite{Engels:2000xw}, \cite{Engels:1999wf}
and \cite{Holtmann:2003he}.

\vskip 3mm
Here we consider the determination of the magnetic equation of state, 
which gives the relation between applied field, temperature and 
magnetization for the system.
The equation of state has been determined perturbatively for general $N$ 
by $\epsilon$-expansions (see \cite[Chapter 29]{Zinn-Justin} and references
therein) and for the cases $N =$ 2, 3, 4 by matching a high-temperature 
expansion (with coefficients obtained from perturbation theory) to a 
parametric form incorporating the leading Goldstone-mode behavior 
\cite{Pelissetto:2000ek,Toldin:2003hq}. Of course it is interesting
to compare these expressions to Monte Carlo data for the equation
of state. One can also {\em test} the various forms used in the 
perturbative expansions (or new proposed forms) by fitting them to 
the Monte Carlo results and obtaining nonperturbative coefficients. 
This has been done (see e.g.\ \cite{Engels:1999wf}) using an 
interpolation of the low-temperature (Goldstone-mode) form derived in
\cite{Wallace:1975vi} with a high-temperature form
determined by analyticity conditions. This method has the advantage
of a clear low-temperature form, with several orders in the 
Goldstone-mode expansion, but has the disadvantage of needing an
interpolation with the high-temperature form.

In the present paper we carry out fits using instead a variant
of Josephson's parame\-tri\-zation \cite{Zinn-Justin,Josephson}, a polynomial 
parametric representation for the equation of state.
The resulting representation is valid above and below 
the critical temperature and automatically satisfies the analyticity 
conditions mentioned above.
In addition to the leading (multiplicative) Goldstone-mode contribution,
we consider explicitly the higher-order terms, which are important
in the low-temperature region. Our proposed form contains two sets
of coefficients, which will be separately more relevant for
the description of the high- or low-temperature regimes. We argue
that the use of this double set of coefficients enables a better 
characterization of the two regimes, leading to better fits
in the comparison with numerical data.
This claim is verified by an application to the $N=4$ case, 
for which the Goldstone-mode effects are fairly high,
using the data reported in \cite{Engels:1999wf}.
As mentioned above, this case is of interest for comparison with data 
from numerical simulations of the phase transition in two-flavor QCD.
In particular, the prediction of universal behavior in the $O(4)$
class has been confirmed
for lattice-QCD data in the Wilson-fermion case 
\cite{Iwasaki:1996ya}, but not for the
staggered-fermion formulation, which is believed
to be the appropriate formulation for studies of the chiral region.
(At the same time, some recent numerical studies suggest that the
transition may be of first order \cite{1storder}.)
We plan to extend our analysis to the $N=2$ case, for which we are 
generating new data \cite{inprep}.

\vskip 3mm
The paper is organized as follows. In Section \ref{eq_state} we describe 
the usual parametric representation for the equation of state, as well as
our proposed form. In Sections \ref{ratios} and \ref{pseudo} we consider
the determination of important universal properties that can be obtained 
from the equation of state: some critical amplitude ratios
and the characterization of the pseudo-critical line (respectively in 
Section \ref{ratios} and in Section \ref{pseudo}).
Finally, in Sections \ref{results} and \ref{conclusion}
we present our results and conclusions.

\section{Scaling equation of state}
\label{eq_state}

The magnetic scaling equation of state is given 
\cite[Chapter 29]{Zinn-Justin} by
\be
h = M^{\delta}\,f(t/M^{1/\beta}),
\ee
where $t$ and $h$ are the reduced temperature $t=(T-T_c)/T_0$ 
and magnetic field $h=H/H_0$.
We fix the normalization constants $T_0$ and $H_0$ by requiring
unit critical amplitudes in the behavior of the magnetization
along the coexistence line (given by $t\to 0_-$, $h = 0$)
and along the critical isotherm (given by $h\to 0$, $t = 0$), corresponding
respectively to $M = (-t)^{\beta}$ and $M = h^{1/\delta}$.

The equation of state can also be written as
\be
y\;=\;f(x)\;,
\label{eqstate}
\ee
where
\be
y \;\equiv\; h/M^{\delta}, \quad x \;\equiv\; t/M^{1/\beta}\;.
\label{xy}
\ee
Note that the coexistence line and the critical isotherm are
given respectively by $x=-1$ and $x=0$.
The corresponding normalization conditions are thus
\be 
f(0) \;=\; 1, \qquad f(-1) \;=\; 0\,. 
\ee

For large values of $x$ (i.e.\ in the high-temperature region of
the phase diagram) the behavior of $f(x)$ is described by Griffiths's 
analyticity condition \cite{Zinn-Justin}
\be
f(x) \;=\; \sum_{n=1}^{\infty} a_n\,x^{\gamma - 2(n-1)\beta} \;.
\label{Griffiths}
\ee

As said in the Introduction,
at low temperatures there appear divergences in the zero-field magnetic 
susceptibility, due to transverse fluctuations from the massless Goldstone 
modes \cite{Zinn-Justin,Parisi,Brezin-Wallace,Lebowitz-Penrose,Lawrie:vk}. 
To leading order 
the divergence in the longitudinal susceptibility is proportional to 
$h^{-1/2}$ and the equation of state has the leading behavior
\be
f(x) \;=\; y \;\propto\; (1+x)^2
\label{lead-goldstone}
\ee
for $x\to -1$.
We note that the Goldstone-mode divergences cancel out and the
equation of state is divergence free. This is seen order by
order in the $\epsilon$-expansion \cite{Brezin-Wallace-Wilson}
and for fixed-dimension perturbation theory \cite{Schaefer-Horner}.
This is also observed nonperturbatively in
the Monte Carlo data (see e.g.\ \cite{Engels:1999wf}).
The corrections to the leading behavior are incorporated explicitly 
in the expression proposed by Wallace and Zia \cite{Wallace:1975vi},
which is inferred from the $\epsilon$-expansion for the equation of
state deduced in \cite{Brezin-Wallace-Wilson}. For $d=3$
the expression corresponds to an expansion in powers of $y^{1/2}$
\be
x_1(y)+1 \;=\; ({\widetilde c_1} \,+\, {\widetilde d_3})\,y \,+\,
             {\widetilde c_2}\,y^{1/2} \,+\,
             {\widetilde d_2}\,y^{3/2} \,+\, \cdots
\label{PTform}
\ee
This form describes well the Monte Carlo data from the low-temperature 
region until around the critical temperature. 
The coefficient associated with the $H^{-1/2}$ divergence of the
susceptibility for $H\to 0$ is ${\widetilde c_2}$. 
Note that the expression of ${\widetilde c_2}$ derived in 
\cite{Wallace:1975vi} increases with $N$, i.e.\ models with larger
$N$ should display stronger Goldstone-mode effects.

In References \cite{Engels:2000xw}, \cite{Engels:1999wf}
and \cite{Holtmann:2003he} the Monte Carlo data for the 
equation of state have been fitted to the expression
\be
x(y) \;=\; x_1(y)\,\frac{y_0^n}{y_0^n + y^n} \,+\,
           x_2(y)\,\frac{y^n}{y_0^n + y^n}~ \,,
\label{totalfit}
\ee
where $x_1(y)$ is given in Eq.\ (\ref{PTform}) above and 
\be
x_2(y)\;=\; a\, y^{1/\gamma} \,+\, b\,y^{(1-2\beta)/\gamma}
\label{highx}
\ee
corresponds to the first two terms of Eq.\ (\ref{Griffiths}).
(The parameters $y_0$ and $n$ are chosen to ensure a good
interpolation.)
This interpolation of low- and high-tempera\-ture behaviors describes
well the data, but of course it would be nicer to have a form valid in both
regions, such as the parametric form introduced in
\cite{Josephson}.
This type of form is described in the next section. We then comment on 
the previous use of this parametrization and propose a new variant that
is especially well suited for fits.

\subsection{Parametric representation}
\label{theta-param}

\begin{figure}[ht]
\vspace{2mm}
\begin{center}
\epsfxsize=0.4\textwidth
\leavevmode\epsffile{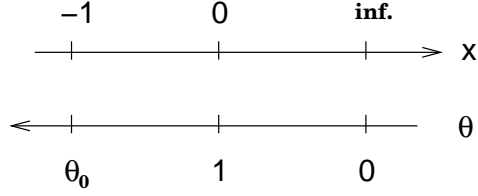}
\caption{\label{xtheta}
Schematic representation of the relation between $x$ and $\theta$. From left
to right, the ticks correspond respectively to the coexistence line, the critical 
point and the high-temperature (zero-field) limit.}
\end{center}
\end{figure}

Let us consider the polynomial parametric representation introduced in
\cite{Josephson}, in which one writes $M$, $t$ and $H$ in terms of 
the variables $R$ and $\theta$ (see e.g.\ \cite{Zinn-Justin,Pelissetto:2000ek})
\ba
&& M \;=\; m_0 \, R^{\beta} \, m(\theta)
\label{eqm}
\\[2mm]
&& t \;=\; R \, \left( 1 - \theta^2 \right)
\label{tofR}
\\[2mm]
&& H \;=\; h_0 \, R^{\beta \delta} \, h(\theta)\;.
\label{eqh}
\ea
Here $m(\theta)$ and $h(\theta)$ are odd functions\footnote{~The function 
$h(\theta)$ should not be confused with $h$, the normalized magnetic field 
introduced in Section \ref{eq_state}.} of $\theta$,
regular at $\theta=0$ and $\theta=1$.
This ensures that Griffiths's analyticity conditions are satisfied.
The coexistence line is given by $\theta_0$, the smallest positive
zero of $h(\theta)$. [From Eq.\ (\ref{tofR}) it is clear that
$\theta_0$ must be greater than 1.]
Without loss of generality, we may take $m(\theta) = \theta$.
The equation of state then becomes
\ba
x &=& {1 - \theta^2\over \theta_0^2 - 1}\;
\left( {\theta_0\over \theta}\right)^{1/\beta}\;,
\label{defx} \\[3mm]
f(x) &=& \theta ^{-\delta} \, {h(\theta)\over h(1)}\;\;.
\label{deffx}
\ea
The relation between $x$ and $\theta$ is shown schematically in Fig.\ \ref{xtheta}
together with the respective ranges considered. Note that we must have 
$\,\theta_0^2 \,<\, 1 / (1-2\beta)\,$
for the above mapping to be invertible \cite{Guida:1996ep}.
For the $O(4)$ case $\beta\sim 0.4$ and we have roughly $\theta_0^2 < 5$.

With the parametrization \reff{eqm}--\reff{eqh}, the
singular part of the free energy ${\cal F}_s$ can be written
as
\be
{\cal F}_s \;=\; h_0 \, m_0 \, R^{2 - \alpha} \, g(\theta)
\, ,
\ee
where the function $g(\theta)$ is the solution of the first-order
differential equation
\be
(1-\theta^2) g'(\theta) + 2(2-\alpha)\theta g(\theta) =
    \left[ (1-\theta^2)m'(\theta) + 2\beta\theta m(\theta) \right]
    h(\theta)
\label{eq:diffeq}
\ee
that is regular at $\theta = 1$ \cite{Zinn-Justin:1999bf}.
(This follows from the relation
$\, H = \partial {\cal F}_s / \partial M\,$,
where the derivative is taken at fixed $t$.)

We note that the parametrization above was first used in perturbative studies of 
the equation of state for the Ising model \cite{Guida:1996ep}.
We discuss below its application to the $N$-vector (Goldstone-mode) case.

In accordance with Eq.\ (\ref{lead-goldstone}) the leading behavior for 
$\theta\to\theta_0$ must be
\be
h(\theta)\;\to\;(\theta_0-\theta)^2 \quad \mbox{for}\;\;  
\theta\,\to\,\theta_0\,.
\ee
This combined with the requirement that $h(\theta)$ be an
expansion in odd powers of $\theta$ suggests the general form
\be
h(\theta) \;=\; \theta \left( 1 - \theta^2/\theta_0^2 \right)^2
        \left( 1 + \sum_{i=1}^n c_{i}\theta^{2i}\right)\;.
\label{hoftheta}
\ee
This form is used in \cite{Pelissetto:2000ek,Toldin:2003hq}, in their
``scheme B''. They also define another scheme with a similar expression
for $m(\theta)$. In both cases the differential equation becomes
\be
(1-\theta^2) g'(\theta) + 2(2-\alpha)\theta g(\theta) =
\theta \, \sum_{i=0}^{3+n} a_i \theta^{2 i}\, ,
\ee
with coefficients $a_i$ depending on the exponent $\beta$, the root $\theta_0$,
the coefficients $c_i$ and on the scheme considered. 
One can easily check that the solution of the differential equation
that is regular at $\theta = 1$ is given by
\be
g(\theta) = - \sum_{i=0}^{3+n} \sum_{k=0}^i 
                  { a_i \over 2 }
             { i! \over \left( i - k \right)! }
             { \theta^{2 (i-k)} \, \left( 1 - \theta^2 \right)^k \over
                     (\alpha - 2) \ldots (\alpha -2 + k)} \, .
\label{gtheta}
\ee
Clearly, this solution is a function of the values of $\theta_0$ 
and of the parameters $c_i$, $i=1,\ldots,n$. 
In \cite{Pelissetto:2000ek,Toldin:2003hq} the authors have 
considered the cases $n=1,2$, with parameters $\theta_0$, $c_i$
obtained from perturbation theory.
We comment on their results in Section \ref{comparison}. 

In the next subsection we introduce a more specific parametric 
expression for $h(\theta)$, as a combined expansion around $\theta=0$
and around $\theta=\theta_0$, in order to isolate the two regions of
the phase diagram.

\subsection{Improved parametric form}
\label{improved_param}
We consider here a variant of the parametric function 
$h(\theta)$ above
\be
h(\theta) \;=\; \theta \left( 1 - \frac{\theta^2}{\theta_0^2} \right)^2
        \left( 1 + \sum_{i=1}^n c_{i}\theta^{2i}\right)\,
        \left[ 1 + {\displaystyle \sum_{j=1}^m} d_{j}
        \left( 1 - \frac{\theta^2}{\theta_0^2} \right)^j  \right] /
        \left( 1 + {\displaystyle \sum_{j=1}^m} d_{j} \right) \,.
\label{h_imp}
\ee
For consistency, the expression is normalized so that the contribution 
from the $d_j$'s is equal to 1 at $\theta=0$.
This normalization factor does not affect the equation of state,
since $h(\theta)$ enters in $f(x)$ only as a ratio.
Let us note that this is still an odd function of $\theta$ and is 
equivalent (as an expansion in $\theta$) to equation (\ref{hoftheta})
with a rearrangement of terms.
In particular, we may compare the series using only terms with $c_i$ 
coefficients (i.e.\ with all $d_j = 0$) with the one using only $d_j$ 
coefficients. The relation between
the two cases is given by
\be
c_i \;\;\Longleftrightarrow\;\;
    \frac{(-1)^i\;{\displaystyle \sum_{j=1}^m} \, 
    \Bigl(\begin{array}{c} j \\ i \end{array}\Bigr)\,d_j}
    {\theta_0^{2i}\;\Bigl( 1\,+\,{\displaystyle \sum_{j=1}^m} d_j\Bigr)}\;.
\ee
Thus, a single coefficient $c_i$ corresponds to a sum of $d_j$'s.
Conversely, if we considered an expansion around 
$\theta\approx\theta_0$ each $d_j$ would correspond to a sum of $c_i$'s.
Since the roles of the terms of the two series are different, considering 
both series may be important for truncated sums, such as the 
ones we use for the fits.

The consideration of two types of coefficients ($c_i$ and $d_j$) is done
for gaining better control over the description of the two distinct regions 
of the phase space, the low- and high-$x$ regions. 
In fact, although the two sets of coefficients give rise to a
similar expansion in powers of $\theta$, the determination of the single
coefficients (by fits to the numerical data) is more stable when each 
of the two regions is separately associated with a set
of coefficients. More precisely, since we write the series as a 
product of two sums of terms corresponding respectively to an expansion 
around the high-$x$ region 
($\theta \approx 0$, coefficients $c_i$) and the low-$x$ region
($\theta \approx \theta_0$, coefficients $d_j$), we can expect
fits of the data for each of these two regions to be more sensitive to 
the corresponding set of coefficients, since the other set's
main contribution will be a constant. An indication of this property
can be seen from a ``quick'' expansion
of the parametric form in powers of $\epsilon \approx 0 $ for the two regions
\begin{eqnarray}
h(\epsilon) &\approx& \epsilon\, 
        \left( 1 - \frac{2\,\epsilon^2}{\theta_0^2} \right)\,
        \left( 1 + c_{1}\epsilon^{2}\right)\,
        \left[ 1 + \sum_{j=1}^m d_{j} 
              \left( 1 - \frac{j\,\epsilon^2}{\theta_0^2} \right) \right] /
        \left( 1 + {\displaystyle \sum_{j=1}^m} d_{j} \right) \qquad
        \label{eq:thetatozero} \\[3mm]
h(\theta_0 - \epsilon) &\approx& \frac{4\,\epsilon^2}{\theta_0}\,
        \left( 1 - \frac{2\,\epsilon}{\theta_0}\right)\,
        \left[ 1 + \sum_{i=1}^n c_{i}\,\theta_0^{2i} \left( 1 -
               \frac{2\,i\,\epsilon}{\theta_0} \right) \right] \;
        \frac{\left(1 + d_{1}\,2\,\epsilon/\theta_0^2\right)}{
        1 + {\displaystyle \sum_{j=1}^m} d_{j}} \,.\qquad
        \label{eq:thetatothetazero}
\end{eqnarray}
Note that we show only the leading order from each multiplicative 
contribution. It is interesting that the correction to the leading behavior 
is ${\cal O}(\epsilon^2)$ in the first case and ${\cal O}(\epsilon)$ in
the second.

To be more precise, we can associate the various $c_i$, $d_j$ coefficients with 
the coefficients in the separate expressions for the low- and high-$x$
regimes used in \cite{Engels:1999wf}, respectively equations (\ref{PTform})
and (\ref{highx}) above.
For the {\bf high-$x$ behavior} we expand $f(x)$
around $\theta\to 0$ (corresponding to large $x$). We start by writing 
$\theta$ as a function of $x$ and inverting Eq.\ (\ref{defx}) consistently in
powers of $\theta$ (correspondingly in powers of $x^{-1/\beta}$). We then get
\begin{eqnarray}
m(\theta) &=& \theta \;\;=\;\; 
          (A\,x)^{-\beta}\,\left[ 1 \,-\,\beta\,(A\,x)^{-2\beta}
                           \,+\,O(x^{-4\beta})\right] \\[2mm]
h(\theta) &=& (A\,x)^{-\beta}\,\left[ 1 \,+\,B\,(A\,x)^{-2\beta}
                           \,+\,O(x^{-4\beta})\right]
\end{eqnarray}
where
\begin{eqnarray}
A &\equiv& (\theta_0^2 \,-\, 1)\, \theta_0^{-1/\beta} \\[3mm]
B &\equiv& c_1 \,-\, \frac{2}{\theta_0^2} \,-\, \beta \,-\,
           \frac{1}{\theta_0^2} \; \frac{\sum_j j\,d_j}
                                   {1 + \sum_j d_j}\,.
\end{eqnarray}
The equation of state then becomes
\begin{equation}
y \;=\; \frac{(A\,x)^{\gamma}}{h(1)} \, \left[ 1 \,+\,
        \left( \beta\,\delta \,+\, B \right)\,(A\,x)^{-2\beta}
                           \,+\,O(x^{-4\beta})\right] \;.
\end{equation}
This expression is of the form (\ref{Griffiths}) and can be inverted and
compared to Eq.\ (\ref{highx}), giving
\begin{eqnarray}
a &=& \frac{1}{A}\,\left[h(1)\right]^{1/\gamma} \\[2mm]
b &=& - \, \frac{a^{1-2\beta}}{\gamma} \,
                         \left(\beta\,\delta \,+\, B \right)\;.
\end{eqnarray}
Note that the leading coefficient $a$ contains $\theta_0$ and sums of
the coefficients $c_i$ and $d_j$, whereas the expression for the
next orders will contain isolated contributions from the $c_i$'s 
(e.g.\ the coefficient $c_1$ in the expression for $b$) but
not from the $d_j$'s, which appear always as a sum.

Analogously, for the {\bf low-$x$ region} we expand the expressions 
of $\theta$, $h(\theta)$ around $\theta\to \theta_0$ (corresponding to 
$x\to -1$) and substitute the results into the expression for $f(x)$.
Defining
\be
\theta \;=\; \theta_0\,(1\,-\,\epsilon)
\ee
we write $x$ as a function of $\epsilon$, invert this expression to get 
$\epsilon(x)$ and then obtain 
$h(\theta)$ in terms of $x$, as done above for the large-$x$ 
case. The expressions are
\begin{eqnarray}
\epsilon &=& 
         \left(\frac{1+x}{A'}\right)
         \,\left[ \,1 \,-\, \frac{B'}{A'}\,
         \left(\frac{1+x}{A'}\right)
         \,+\,\left(\frac{2{B'}^2}{{A'}^2}\,-\,
         \frac{C}{A'}\right)\,
         \left(\frac{1+x}{A'}\right)^2
         \,+\,\cdots\,\right] \\[3mm]
h(\theta) &=& D\,\left(\frac{1+x}{A'}\right)^2\,
         \left[ \,1 \,+\, \left(E\,-\,\frac{2B'}{A'}\right)\,
         \left(\frac{1+x}{A'}\right) \right. \nonumber \\[2mm]
         & & \left. \hskip 25mm
         \,+\, \left(F\,-\,\frac{3B'E+2C}{A'} \,+\, \frac{5{B'}^2}{A'^2} 
         \right)\,
         \left(\frac{1+x}{A'}\right)^2 \,+\, \cdots \,\right]
\end{eqnarray}
with
\begin{eqnarray}
A' &\equiv& \frac{2\theta_0^2}{\theta_0^2-1} \,-\, 
     \frac{1}{\beta} \\[2mm]
B' &\equiv& -\,\frac{\theta_0^2}{\theta_0^2-1}\,
               \left(1\,-\,\frac{2}{\beta} \right) \,-\, 
    \frac{1}{2\beta} \,\left(\frac{1}{\beta}+1\right) \\[2mm]
C &\equiv&  -\, \frac{1}{6\beta} \,\left(\frac{1}{\beta}+1\right)
            \left(\frac{1}{\beta}+2\right) \,+\,
            \frac{\theta_0^2}{\theta_0^2-1}\,\frac{1}{\beta^2}\\[2mm]
D  &\equiv& 4\,\theta_0\,\left(1\,+\,\sum_i c_i\,\theta_0^{2i}\right) /
            \left( 1 + \sum_j d_j \right) \\[2mm]
E  &\equiv& 2 d_1 \,-\, 2 \,-\, \frac{\sum_i 2i\, c_i \,\theta_0^{2i}}
            {1\,+\,\sum_i c_i\,\theta_0^{2i}}\\[3mm]
F  &\equiv&  4 d_2 \,-\, 5 d_1 \,+\, \frac{5}{4} \,+\,
            \frac{4\,(1-d_1)\,\sum_i i\, c_i \,\theta_0^{2i}}
            {1\,+\,\sum_i c_i\,\theta_0^{2i}} \,+\,
            \frac{\sum_i i\,(2i-1)\, c_i \,\theta_0^{2i}}
            {1\,+\,\sum_i c_i\,\theta_0^{2i}} \;. 
\end{eqnarray}
The equation of state then becomes
\be
y \;=\; \frac{D\,\theta_0^{-\delta}}{h(1)}\,
   \left(\frac{1+x}{A'}\right)^2 \,\left[\,1\,+\,
   \left(E-\frac{2B'}{A'}+\delta\right)\,\left(\frac{1+x}{A'}\right)
   \,+\, G\,\left(\frac{1+x}{A'}\right)^2\,+\,
   \cdots\,\right]\;,
\ee
where
\be
G \;\;\equiv\;\; \frac{\delta (\delta+1)}{2}\,+\,\delta\,E \,+\, F
             \,-\,\frac{3\delta \,B'\,+\, 3B'E \,+\, 2C}{A'}
             \,+\, \frac{5{B'}^2}{{A'}^2}\;.
\ee
This form may be inverted to give an expression of $x$ as a series of
powers of $y^{1/2}$ as in Eq.\ (\ref{PTform}). We obtain the coefficients
\begin{eqnarray}   
{\widetilde c_2} &=& A'\,
       \left[\frac{\theta_0^{\delta}\,h(1)}{D}\right]^{1/2} \\[2mm]
{\widetilde c_1} \,+\, {\widetilde d_3} &=& 
                     -\,\frac{{\widetilde c_2}^2}{2A'}\,
                     \left( E\,-\,\frac{2B'}{A'}\,+\,\delta\right) \\[2mm]
{\widetilde d_2} &=& \frac{{\widetilde c_2}^3}{2{A'}^2}\,
                     \left[\,\frac{5}{4}\,
                     \left(E\,-\,\frac{2B'}{A'}\,+\,\delta\right)^2 
                     \,-\,G \,\right] \;.
\end{eqnarray}
We see that in this case it is the $d_j$ coefficients that appear
as single contributions, whereas the $c_i$'s appear always as sums.

Thus, the qualitative feature observed in Eqs.\ (\ref{eq:thetatozero}) and
(\ref{eq:thetatothetazero}) is confirmed by a more careful expansion, i.e.\ 
the $c$'s are more relevant for the high-$x$ region and vice-versa.
This will also be seen directly from the fits in Section \ref{fits}.
All calculations above were checked using {\tt Mathematica}.

\section{Amplitude ratios}
\label{ratios}
Just like other critical properties of statistical systems
(e.g.\ critical exponents), certain ratios of critical amplitudes
are universal \cite{Privman}. 
The amplitude ratios are taken as dimensionless combinations
of critical amplitudes above and below $T_c$ for various quantities.
For example, for the singular part of the specific heat one has
\be
C_H \;=\; A^{\pm}\,\left\vert t \right\vert^{-\alpha},\qquad 
t \rightarrow \pm 0\,,
\ee
where $t\,\propto\,(T-T_c)$.
The ratio $A^+/A^-$ is then universal.
Similarly, by considering the behaviors of 
\begin{itemize}
\item the susceptibility along the critical isochore ($t>0$, $H=0$)
\be
\chi \;=\; C^{+} t^{-\gamma}
\ee
\item the magnetization along the critical isotherm ($t=0,H\neq0$)
\be
M \;=\; D_c^{-1/\delta} H^{1/\delta}
\ee
\item the magnetization on the coexistence line ($t<0$, $H=0$)
\be
M \;=\; B (-t)^{\beta}
\ee
\end{itemize}
one may construct the universal ratios
\ba
R_c &=& \alpha\,A^+ C^+ / B^2 \;,\\
R_{\chi} &=& C^+ D_c\, B^{\delta-1} \;.
\ea
These and other universal ratios may be obtained directly from Monte Carlo
simulations (as done e.g.\ in \cite{Cucchieri:2002hu}) or indirectly from 
the equation of state.
In the case of the Josephson parametrization discussed above, the
universal amplitude ratios of quantities defined at zero momentum are
given in terms of $g(\theta)$ by \cite{Guida:1996ep}
\begin{eqnarray}
&& A^+/A^- \;=\;
(\theta_0^2 - 1 )^{2-\alpha} \, {g(0)\over g(\theta_0)}\,,
 \\[2mm]
&& R_c \;=\; - \,\alpha \,
   (1-\alpha)\,(2-\alpha) \, \frac{(\theta_0^2 - 1 )^{2\beta} \,g(0)}{
      \theta_0^2 \, h'(0)} \,,\\[2mm]
&& R_\chi \;=\; \frac{\theta_0^{\delta-1}\,h(1)}{
                (\theta_0^2 - 1 )^{\gamma}\,h'(0)} \,.
\end{eqnarray}
Note that in order to evaluate the function $g(\theta)$ one has to
solve the differential equation (\ref{eq:diffeq}), i.e.\
determine the solution (\ref{gtheta}).

Our results for the above ratios are reported in Section \ref{universal}.

\section{The pseudo-critical line}
\label{pseudo}

Another important property that can be extracted from the equation
of state is the characterization of the so-called pseudo-critical line,
defined by the points where
the susceptibility $\chi$ shows a (finite) peak for $H\neq 0$. This
corresponds to the rounding of the divergence observed
at the critical point, i.e.\ for $H=0$ and $\,T=T_c$.
More precisely, one looks for a peak in the scaling function of the
susceptibility, given by \cite{Engels:2001bq}
\be
M\;=\;h^{1/\delta}\,f_M(z) \;\;\Rightarrow\;\;
\chi \;=\; \frac{\partial M}{\partial H} \;=\; 
\frac{h^{1/\delta - 1}}{H_0}\,f_{\chi}(z)\;,
\ee
where 
\be
z\;\equiv\; t/h^{1/{\beta\delta}}\;.
\ee
Clearly, at each fixed $h$ the peak in $\chi$ is given by
$\,t_{p}\,=\, z_p\,h^{1/\beta \delta}$,
and we have
\be
M_p\,=\, h^{1/\delta}\,f_M(z_p), \;\quad\;
H_0\,\chi_{p}\,=\, h^{1/\delta - 1} \,
f_{\chi}(z_p)\,.
\ee
Thus, the behavior along the pseudo-critical line is determined by the
universal constants $z_p$, $\,f_M(z_p)$, $\,f_{\chi}(z_p)$. Determining
this line is important for systems where a study at $H=0$ is not
possible (and consequently the critical value $T_c$ is not known
with accuracy), such as for the chiral transition of QCD at finite temperature.
In fact, the knowledge of these universal constants allows an
unambiguous normalization of QCD data (using the observed scaling along the
pseudo-critical line), as done in \cite{unamb}.

The pseudo-critical line has been studied for $O(2)$ and $O(4)$ models in 
\cite{Engels:2001bq,Schulze:2001zg}. For the $N=4$ case,
it is found that the susceptibility peaks are given by $z_p = 1.33(5)$.
Since this value is close to the interpolating point of
the equation of state in \cite{Engels:1999wf}, 
it is very important to work with the smooth parametrization 
considered here, especially when using the derivative of $f_M(z)$ as
in Eq.\ (\ref{eq:fchi}) below.

The expression for $f_{\chi}(z)$ can be easily obtained from 
the equation of state, given by $f_M(z)$ or $f(x)$. 
Using the original parametrization we obtain \cite{Engels:2001bq}
\be
f_{\chi}(z) \;=\; \frac{1}{\delta}\,
\left[f_M(z)\,-\,\frac{z}{\beta}\,f_M'(z)\right]
\;=\; \frac{\beta\,[f(x)]^{1-1/\delta}}{\beta\,\delta\, f(x) \,-\, x\,f'(x)}\;.
\label{eq:fchi}
\ee
(Note that $\,z\,=\,x\,[f(x)]^{-1/\beta\delta}$.)
In terms of the parametric representation this gives
\begin{eqnarray}
f_{\chi}(\theta) &=& \left[\frac{h(\theta)}{h(1)}\right]^{-1/\delta}\,
         \frac{(2\,\beta\,\theta^2\,+\,1\,-\,\theta^2)\,h(\theta)}{
               2\,\beta\,\delta\,\theta\,h(\theta) \,+\, 
              (1\,-\,\theta^2)\,h'(\theta)} 
\label{eq:fchi_theta}
\\[2mm]
z(\theta) &=& \left[\frac{h(\theta)}{h(1)}\right]^{-1/\beta\delta}\,
         \frac{\theta_0^{1/\beta}\,(1-\theta^2)}{\theta_0^2 - 1}\;.
\label{eq:z_theta}
\end{eqnarray}

Our results for $f_{\chi}(z)$ and the determination of $z_p$ are shown 
in Section \ref{universal}.

\section{Results}
\label{results}

The fits have been done using a conjugate-gradient minimization
\cite{Numerical_recipes}
of $\chi^2$ --- without considering the gradient of the function 
$f(\theta)$ --- 
with a numerical inversion of Eq.\ (\ref{defx}) in order to find $\theta$
for any given value of $x$.
For the critical exponents we used $\nu = 0.749(2)$
\cite{Hasenbusch:2000ph}
and $\delta = 4.824(9)$ \cite{Engels:2003nq}, implying the
values $\beta = 0.386(1)$, $\gamma = 1.476(5)$ and the upper bound 
$\theta_0^2 \leq 4.38(5)$.
We refer to these values as the {\em first} set of exponents.
We note that the corresponding exponent $\delta$ from 
\cite{Hasenbusch:2000ph}, $4.789(6)$, has slightly smaller error bars.
However, we choose to use the one in \cite{Engels:2003nq} because it
is obtained directly from (infinite-volume) simulations at nonzero magnetic 
field. These two exponents are {\em not} in agreement within 
error bars. We will also present below for comparison a few quantities 
obtained using the exponent $\delta$ from \cite{Hasenbusch:2000ph}.
We refer to the resulting values as the {\em second} set of critical
exponents.

The data for the magnetization are taken from \cite{Engels:1999wf}.
In addition to the statistical errors, we have included errors due to
the critical exponents, the critical temperature and the normalization
constants $H_0$ and $T_0$. These constants have been rederived using
the first set of exponents above (with errors), yielding
\be
H_0 \;=\; 4.85(2)\;,\quad \; T_0 \;=\; 1.055(5)\;.
\ee
The errors reported in parentheses
in all the tables below are Monte Carlo (MC) errors, obtained with
2000 MC iterations. In particular, in Section \ref{fits} we not only vary
the $y$ variable but also consider the uncertainties in the exponents 
$\gamma$ and $\delta$ appearing in the fitting function [i.e.\ in Eqs.\ 
(\ref{defx}) and (\ref{deffx})]. The same is true for the errors reported
in Section \ref{comparison}. In Section \ref{universal} the errors
comprise the error bars in the input parameters and also
the errors in the critical exponents.

\begin{table}[t]
\begin{center}
\begin{tabular}{|c|c|c|c|} \hline
$\theta_0^2$ & $c_1$ & $c_2$ & $\chi^2 / d.o.f.$ \\ \hline
  2.33(3)    &       &       &    0.50 \\ \hline
  2.01(8)    & 0.16(6)  &       &    0.43 \\ \hline
  1.67(5)    & 0.22(6) & 0.18(4) &    0.44 \\ \hline \hline
$\theta_0^2$ & $d_1$ & $d_2$ & $\chi^2 / d.o.f.$ \\ \hline
  3.61(4)    & 438(1)  &       &    0.59 \\ \hline
  7(2)       & -1.4(2)   & -0.075(1) & 0.44 \\ \hline
\end{tabular}
\end{center}
\caption{Fits in the high-temperature regime (using $x\geq 0$). The values of
$\chi^2 / d.o.f.$ should be taken only as relative measures of the goodness
of the fits. The number of $d.o.f.$\ is 33.}
\vskip 1cm
\label{table:HT}
\end{table}
\begin{table}[ht]
\begin{center}
\begin{tabular}{|c|c|c|c|} \hline
$\theta_0^2$ & $c_1$ & $c_2$ & $\chi^2 / d.o.f.$ \\ \hline
  1.905(7)   &       &       &  40.1   \\ \hline
  1.09(2)    & -1.18(4) &       &  27.3   \\ \hline
  1.07(1)    & 3.3(1) & -5.1(1) & 28.1    \\ \hline \hline
$\theta_0^2$ & $d_1$ & $d_2$ & $\chi^2 / d.o.f.$ \\ \hline
  3.85(4)    & -4.0(2) &       &   20.2  \\ \hline
  2.69(2)    & 154(4)  & -111(2) & 18.7 \\ \hline
\end{tabular}
\end{center}
\caption{Fits in the low-temperature regime (using $x\leq 0$). The values of
$\chi^2 / d.o.f.$ should be taken only as relative measures of the goodness     
of the fits. The number of $d.o.f.$\ is 37.}
\vskip 1cm
\label{table:LT}
\end{table}

\subsection{Fits}
\label{fits}
\begin{table}[ht]
\vspace{1cm}
\begin{center}
\begin{tabular}{|c|c|c|c|c|c|} \hline
$\theta_0^2$ & $c_1$ & $c_2$ & $c_3$ & $c_4$ & $\chi^2 / d.o.f.$ \\ \hline
   1.955(7)   &             &          &           &          &  31.5 \\ \hline
   1.614(7)   &    0.58(3)  &          &           &          &  19.6 \\ \hline
   1.392(5)   & -0.06(1)    & 0.80(3)  &           &          &  18.1 \\ \hline
   1.247(6)   & 1.6(2)      & -2.8(3)  & 2.7(2)    &          &  17.6 \\ \hline
   1.170(3)   & -0.7(1)     & 6.8(4)   & -11.6(8)  & 7.2(5)   &  17.4 \\ \hline
\end{tabular}
\end{center}
\caption{Fits using only $c_i$ terms and the whole set of data. The values of
$\chi^2 / d.o.f.$ should be taken only as relative measures of the goodness     
of the fits. The number of $d.o.f.$\ is 69.}
\vskip 1cm
\label{table:onlyc}
\end{table}
\begin{table}[ht]
\begin{center}
\begin{tabular}{|c|c|c|c|c|c|} \hline
$\theta_0^2$ & $d_1$ & $d_2$ & $d_3$ & $d_4$ & $\chi^2 / d.o.f.$ \\ \hline
   3.99(4)    & -3.5(1)  &       &       &       &  12.0   \\ \hline
   3.22(4)    & -9.2(9)  & 3.9(5)   &       &       &  10.2   \\ \hline
   2.63(2)    & -69(3)   & 83(2)    & -36(1)  &       &  10.0   \\ \hline
   2.73(2)    & -53(3)   & 42(2)    & 5.9(2)  & -15.0(9)  &  10.2   \\ \hline
\end{tabular}
\end{center}
\caption{Fits using only $d_j$ terms and the whole set of data.  The values of
$\chi^2 / d.o.f.$ should be taken only as relative measures of the goodness     
of the fits. The number of $d.o.f.$\ is 69.}
\vskip 1cm
\label{table:onlyd}
\end{table}
\begin{table}[ht]
\footnotesize
\begin{center}
\begin{tabular}{|c|c|c|c|c|c|c|} \hline
fit ($c_i + d_j$) &
  $\theta_0^2$ & $c_1$ & $c_2$ & $c_3$ & $d_1$ & $\chi^2 / d.o.f.$ \\ \hline
3+1 & 2.16(3)  & 0.80(6) & -0.39(7)  & 0.58(4)   & 33(6)    & 9.8 \\ \hline
  & $\theta_0^2$ & $c_1$ & $c_2$ & $d_1$ & $d_2$ & $\chi^2 / d.o.f.$ \\ \hline
2+2 & 2.17(4)   & 0.9(1)   & -0.62(7)  & -1.56(4)& 1.15(5)&  9.8  \\ \hline
  & $\theta_0^2$ & $c_1$ & $d_1$ & $d_2$ & $d_3$ & $\chi^2 / d.o.f.$ \\ \hline
1+3 & 2.16(2)   & 1.4(1)   & 31.2(9)   & -50(1)   & 38(2)   & 9.8 \\ \hline
\end{tabular}
\end{center}
\caption{Fits using 5 parameters and the whole set of data. The 
number of coefficients $c_i$ and $d_j$ used in each case is
indicated in the first column. The values of     
$\chi^2 / d.o.f.$ should be taken only as relative measures of the goodness     
of the fits. Here we use the first set of critical exponents.
(The number of $d.o.f.$\ is 69.)}
\vskip 1cm
\label{table:good}
\end{table}
\normalsize

As a first step, we tried to fit the data separately in the high- and 
low-$x$ regimes using only $c_i$ or only $d_j$ parameters, in order 
to confirm that the $c_i$'s are more important at high $x$
and the $d_j$'s at low $x$, as suggested in Section \ref{improved_param}. 
As one can see from Tables \ref{table:HT} and \ref{table:LT}, this is 
indeed the case.
At high $x$ the fits using $c_i$ parameters
work better that the fits using $d_j$ parameters,
as can be seen in the case with one parameter
plus $\theta_0^2$. When using two parameters plus $\theta_0^2$ the 
values of $\chi^2 / d.o.f.$ obtained in the two cases
coincide, but in the case with $d_1$ and $d_2$ one obtains
the unphysical value $\theta_0^2 \approx 7$. Moreover, if one
tries to do a fit using $\theta_0^2, \, c_1,\, c_2\,$ and
$d_1$ the fit is not better than the one reported in
the third row of Table \ref{table:HT} and the value of $d_1$ is very
close to 0. In the low-$x$
region the fits using $d_j$ parameters work much better than
the corresponding fits using $c_i$ parameters. Again, if one
tries a fit using $\theta_0^2, \, c_1,\, d_1\,$ and
$d_2$ the result is not better than the one reported
in the last row of Table \ref{table:LT}.
Thus, we see clearly that the coefficients $c_i$ and $d_j$ are more
relevant respectively at high and low $x$, as suggested in
Section \ref{improved_param}.
\begin{figure}[tb]
\begin{center}
\vspace{-40mm}
\epsfxsize=0.5\textwidth
\leavevmode\epsffile{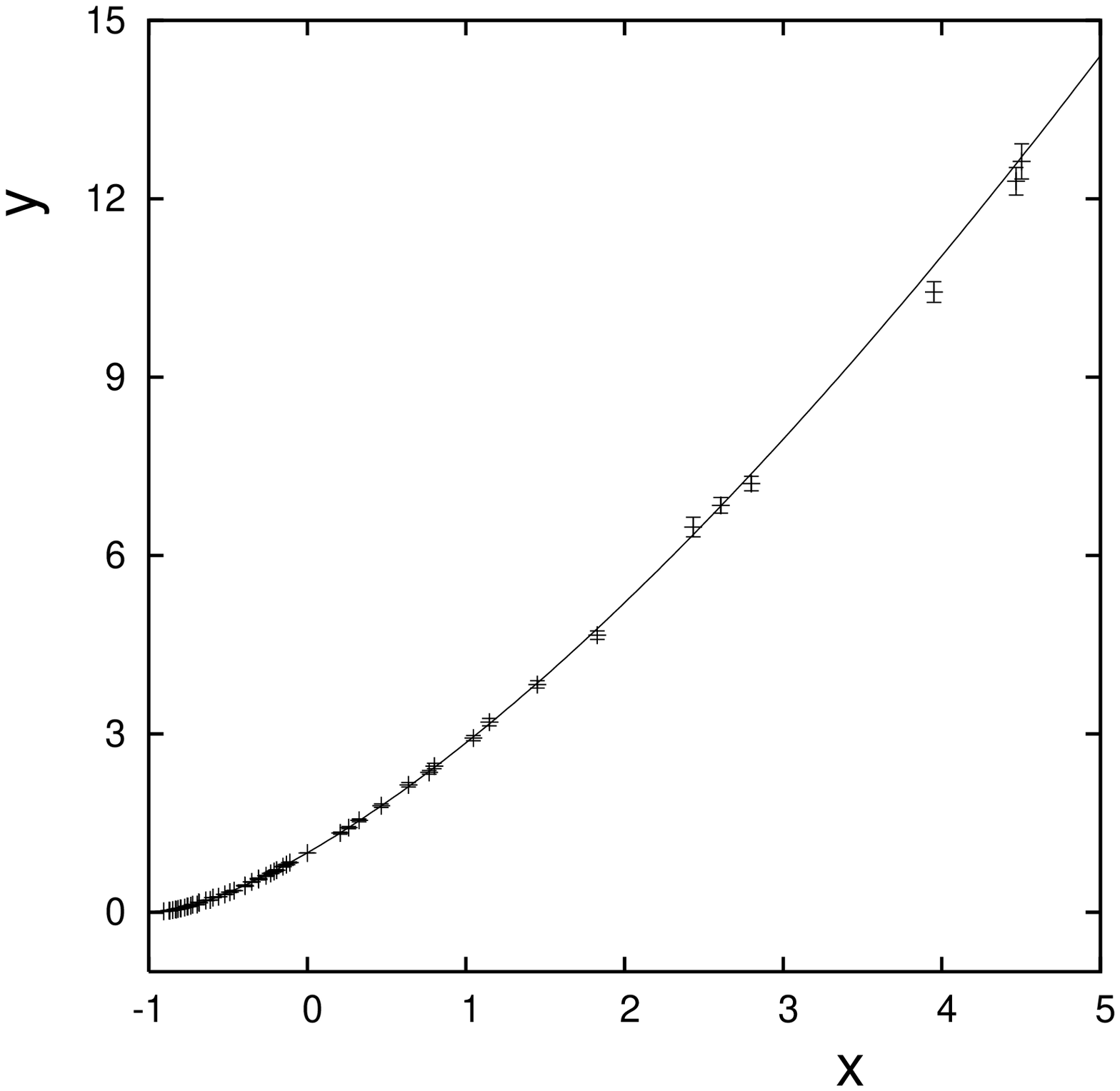}
\caption{\label{fig}
Plot of the data together with the fitting curve for the case
with coefficients $c_1$, $c_2$ and $d_1$, $d_2$ using the first
set of exponents. No errors are shown for the curve.
Error bars on the data are one standard deviation.}
\end{center}
\end{figure}
\begin{table}[ht]
\footnotesize
\begin{center}
\begin{tabular}{|c|c|c|c|c|c|c|} \hline
fit ($c_i + d_j$) &
  $\theta_0^2$ & $c_1$ & $c_2$ & $c_3$ & $c_4$ & $\chi^2 / d.o.f.$ \\ \hline
4+0 & 1.137(2) & -1.0(2) & 8.5(6) & -14.6(9) & 9.0(5) & 58.1 \\ \hline
  & $\theta_0^2$ & $c_1$ & $c_2$ & $c_3$ & $d_1$ & $\chi^2 / d.o.f.$ \\ \hline
3+1 & 2.054(7) &  1.17(6) & -1.05(6) & 1.11(4) & 510(30) & 27.0 \\ \hline
  & $\theta_0^2$ & $c_1$ & $c_2$ & $d_1$ & $d_2$ & $\chi^2 / d.o.f.$ \\ \hline
2+2 & 2.19(3)  & 1.2(1) & -0.80(6) & -1.74(3) & 1.34(4) & 26.6 \\ \hline
  & $\theta_0^2$ & $c_1$ & $d_1$ & $d_2$ & $d_3$ & $\chi^2 / d.o.f.$ \\ \hline
1+3 & 2.22(1) & 0.67(1) & 630(20) & -920(20) & 560(10) & 29.0 \\ \hline
  & $\theta_0^2$ & $d_1$ & $d_2$ & $d_3$ & $d_4$ & $\chi^2 / d.o.f.$ \\ \hline
0+4 & 2.13(1) & -71(2) & 180(4)   & -193(3)  & 75(2)   & 26.7 \\ \hline
\end{tabular}
\end{center}
\caption{Fits using 5 parameters and the whole set of data. The
number of coefficients $c_i$ and $d_j$ used in each case is
indicated in the first column. The values of
$\chi^2 / d.o.f.$ should be taken only as relative measures of the goodness
of the fits. Here we use the second set of critical exponents.
(The number of $d.o.f.$\ is 69.)}
\vskip 1cm
\label{table:other}
\end{table}
\normalsize

As a second step, we checked that the fit of all the data using
only the parameters $c_i$ does not work very well 
(see Table \ref{table:onlyc}).
In particular, even with four parameters $c_i$ one cannot get a large
improvement in the value of $\chi^2 / d.o.f.$, compared to the
case with only the parameter $c_1$. The situation is slightly better
when considering only $d_j$ parameters (see Table \ref{table:onlyd}).
Notice, however, that we have only a few data points
with very large $x$ and that the low-$x$ expression
used in \cite{Engels:1999wf} describes well the data up to 
$x \approx 2$.

Finally, fits of all the data with both $c_i$ and $d_j$ parameters 
(see Table \ref{table:good})
work very well, giving a value of $\chi^2 / d.o.f.$ about
a factor two smaller than the best result obtained in
Table \ref{table:onlyc} (see last row). In particular, we find it
interesting that for the three fits considered
we get (within errors) the same value for $\theta_0^2$. By averaging over
the three results we find 
\begin{equation}
\theta_0^2 \;=\; 2.16(2) \;.
\end{equation}
In Fig.\ \ref{fig} we show a plot of the data together
with the curve corresponding 
to the case on the second row of Table \ref{table:good}.
We have also tried fits with 6 parameters, without significant improvement
in the value of $\chi^2 / d.o.f.$

The relatively high values of $\chi^2 / d.o.f.$ may be related to 
remaining systematic effects in the data.
This is especially true in the low-temperature
regime, where the finite-size effects are very strong due to the effect of
Goldstone-mode-induced singularities. It would be interesting to
test our parametrization using the higher-precision data recently 
produced in \cite{Engels:2003nq}.
In any case, if we use the second set of exponents above
(i.e.\ with $\delta$ from \cite{Hasenbusch:2000ph})
the values of $\chi^2 / d.o.f.$ are significantly worse, as can be seen 
in Table \ref{table:other}.
We note that, in order to consider this second set of exponents, 
we have reevaluated the normalization constants $H_0$ and $T_0$,
the values of $x$ and $y$ and the data errors for this case.

\subsection{Comparison with other parametrizations}
\label{comparison}

We now compare our results with previous expressions for the $O(4)$
equation of state.
In Reference \cite{Toldin:2003hq}, the scheme B considered by the authors 
corresponds to all $d_j = 0$ and only $\theta_0$, $c_1$ nonzero.
Their values for these coefficients are
\be
\theta_0^2 \;=\; 2.4(2) \;, \quad\; c_1\;=\; 0.065(30)\;.
\ee
Note that their value of $\theta_0^2$ is consistent with ours within
error bars.
Using these coefficients as a ``fit'' of the data (considering the first
set of critical exponents above), one obtains a
$\chi^2 / d.o.f.$ of 268. We can also use our second set of data
to evaluate $\chi^2 / d.o.f.$, but this yields the value 688.

We also consider the interpolated parametrization introduced in
\cite{Engels:1999wf}, presented in Eqs.\ (\ref{PTform}),
(\ref{totalfit}) and (\ref{highx}) above. 
Using the first set of data, we obtain the high-$x$ coefficients
\begin{equation}
a \;=\; 1.07(1)\,, \;\;\qquad b \;=\; -0.95(3)
\label{eq:a}
\end{equation}
with $\chi^2 / d.o.f. = 0.52$ (cut at $x = 1.5$). At low $x$ we get
\begin{equation}
{\widetilde c_1} \,+\, {\widetilde d_3} = 0.19(1)\,, \;\;\qquad
{\widetilde c_2} = 0.746(3)\,, \;\;\qquad
{\widetilde d_2} = 0.061(8)
\end{equation}
with $\chi^2 / d.o.f. = 25.5$.
Note that the above coefficients are only in partial agreement with 
the values in \cite{Engels:1999wf} and \cite{Engels:2003nq}, 
mostly due to the slightly different critical exponents considered.
We then use these coefficients for the interpolated expression
in Eq.\ (\ref{totalfit}), setting (as in \cite{Engels:1999wf})
$y_0=10$, $n=3$. The resulting
5-parameter fit of the data has $\chi^2 / d.o.f. = 26.2$.

\subsection{Universal quantities}
\label{universal}

As discussed in Sections \ref{ratios} and \ref{pseudo}, we use the 
fits obtained above to evaluate several interesting universal 
quantities, such as critical amplitude ratios and the characterization
of the pseudo-critical line in the phase diagram.

\begin{table}[ht]
\begin{center}
\vspace{8mm}
\begin{tabular}{|c|c|c|c|} \hline
fit ($c_i + d_j$) &
  $A^+/A^-$ & $R_c$ & $R_\chi$ \\ \hline
3+1 & 1.7(2) & 0.26(2) & 1.11(6) \\ \hline
2+2 & 1.8(5) & 0.26(2) & 1.1(1) \\ \hline
1+3 & 1.8(4) & 0.26(2) & 1.1(1) \\ \hline
\end{tabular}
\end{center}
\caption{Results for the universal amplitude ratios
using the fits reported in Table
\protect\ref{table:good}.}
\vskip 1cm 
\label{table:ratios}
\end{table}

\begin{table}[ht]
\begin{center}
\begin{tabular}{|c|c|c|c|} \hline
fit ($c_i + d_j$) &
  $A^+/A^-$ & $R_c$ & $R_\chi$ \\ \hline
3+1 & 1.6(1) & 0.25(1) & 1.09(5) \\ \hline
2+2 & 1.6(1) & 0.27(2) & 1.1(1) \\ \hline
1+3 & 1.9(3) & 0.22(1) & 1.02(5) \\ \hline
\end{tabular}
\end{center}
\caption{Results for the universal amplitude ratios
using fits reported in Table
\protect\ref{table:other}.}
\vskip 1cm 
\label{table:ratios2}
\end{table}

In Table \ref{table:ratios} we show the results obtained for
the ratios $A^+/A^-$, $R_c$, $R_\chi$ using our preferred fits
(reported in Table \ref{table:good}). The three fits give 
consistent results within error bars. Averaging over the three
cases yields
\begin{equation}
A^+/A^- \;=\; 1.8(2)\,, \;\;\qquad R_c \;=\; 0.26(1)\,, \;\;\qquad 
R_\chi \;=\; 1.10(5)\;.
\label{eq:ratio_results}
\end{equation}
These values are in agreement with the ones reported in 
\cite[Table 3]{Toldin:2003hq}.
(Note, however, that our values take into account the errors due
to the uncertainty in the critical exponents.)
We also show, in Table \ref{table:ratios2}, the same quantities using 
the fits for our second set of data (from Table \ref{table:other}).
These results show a little more fluctuation, but are essentially
in agreement with the ones in Eq.\ (\ref{eq:ratio_results}) above.
Note that the ratio $R_\chi$ can also be evaluated directly
\cite{Engels:2000xw}
from the coefficient $a$ in the interpolated form, given in Eq.\
(\ref{eq:a}). In this case we get $R_\chi = a^\gamma = 1.105(15)$,
in agreement with our result above and with Ref.\ \cite{Engels:2003nq}.

\begin{figure}[tb]
\begin{center}
\vspace{-40mm}
\epsfxsize=0.5\textwidth
\leavevmode\epsffile{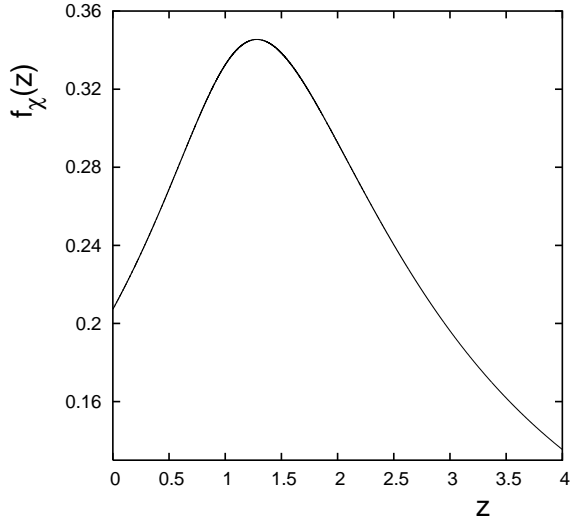}
\caption{\label{fig:pseudo}
Plot of the scaling function of the susceptibility
$f_\chi(z)$ from Eqs.\ (\ref{eq:fchi_theta}) and (\ref{eq:z_theta}),
using the fit in the second row of Table \protect\ref{table:good}.}
\end{center}
\end{figure}

\begin{table}[t]
\begin{center}
\begin{tabular}{|c|c|c|c|} \hline 
fit ($c_i + d_j$) &
  $\theta_p$ & $z_p$ & $f_\chi(z_p)$ \\ \hline
3+1 & 0.580(2) & 1.33(1) & 0.340(2)  \\ \hline
2+2 & 0.589(4) & 1.28(3) & 0.343(3)  \\ \hline
1+3 & 0.592(3) & 1.27(1) & 0.339(2)  \\ \hline
\end{tabular}
\end{center}
\caption{Results for $\theta_p$, $z_p$ and
$f_{\chi}(z)$ using the fits reported in Table
\protect\ref{table:good}.}
\vskip 1cm
\label{table:pseudo}
\end{table}
\begin{table}[ht]
\begin{center}
\begin{tabular}{|c|c|c|c|} \hline
fit ($c_i + d_j$) &
  $\theta_p$ & $z_p$ & $f_\chi(z_p)$ \\ \hline
3+1 & 0.605(2) & 1.25(1) & 0.345(2)  \\ \hline
2+2 & 0.604(3) & 1.21(3) & 0.344(2)  \\ \hline
1+3 & 0.557(3) & 1.33(2) & 0.3553(7)  \\ \hline
\end{tabular}
\end{center}
\caption{Results for $\theta_p$, $z_p$ and
$f_{\chi}(z)$ using fits reported in Table
\protect\ref{table:other}.}
\vskip 1cm
\label{table:pseudo2}
\end{table}

\vskip 3mm
We now turn to the numerical characterization of the pseudo-critical
line (see Section \ref{pseudo}).
Using Eqs.\ (\ref{eq:fchi_theta}) and (\ref{eq:z_theta}),
we draw the parametric plot of the scaling function 
for the susceptibility versus $z$ (see Fig.\ \ref{fig:pseudo}).
The peak corresponds to the pseudo-critical line and can be
determined numerically from the two equations by varying $\theta$.
The peak coordinates thus obtained are reported in Table 
\ref{table:pseudo}, where we used our preferred fits. 
The values are consistent within errors, yielding
\begin{equation}
\theta_p \;=\; 0.587(2)\,, \;\;\qquad
z_p \;=\; 1.29(1)\,, \;\;\qquad f_\chi(z_p) \;=\; 0.341(1)\;.
\end{equation}
The results are in agreement with previous determinations of
$z_p$ and $f_\chi(z_p)$, made in Refs.\ \cite{Engels:2001bq}
and \cite{Toldin:2003hq}, but our error for $z_p$ is
much smaller. In Table \ref{table:pseudo2} we present these
quantities in the case of our second set of data. Again, the 
determinations are in agreement.

\section{Conclusions}
\label{conclusion}

We have introduced an improved parametric form for the description
of the equation of state of $3d$ $O(N)$ models. This form is based
on the parametrization used perturbatively in \cite{Guida:1996ep} 
for the Ising model, 
but takes into account terms associated with the effects of
Goldstone-mode fluctuations. Such effects are present in $O(N)$ 
models along the coexistence line, i.e.\ at low temperatures and 
small magnetic field (or equivalently, 
at low values of the variable $x$). These new
terms are included by means of the $d_j$ coefficients, 
associated with an expansion around the coexistence line.
(The $d_j$'s are considered in addition to the usual $c_i$ 
coefficients, related to the high-temperature/high-$x$ behavior.) 
We show that the new parametric form indeed provides 
a better fit to the numerical data as compared to previous 
parametrizations. In particular, the consideration of the $d_j$ 
coefficients is essential for a good description of the 
Monte Carlo data in the whole range of values of $x$.
Also, we were able to verify clearly the different roles played
by $c_i$ and $d_j$ parameters in the high- and low-$x$ regions.

We note that --- in the case where all $d_j = 0$ --- our parametrization 
is equivalent to the scheme B discussed in \cite{Pelissetto:2000ek}, 
used perturbatively by the authors for general $O(N)$ models.
We find that our value of $\theta_0$ is consistent with their
perturbative determination for the $O(4)$ case, presented in 
\cite{Toldin:2003hq}. 
However, we do not confirm their conjecture that the $c_i$'s get smaller 
with increasing $i$.

We also stress that,
in addition to providing a better fit to the numerical data, the
expression considered is a continuous function, needing no
interpolation between the two $x$ regions. This is particularly
useful for the determination of the pseudo-critical line, since
the interpolating form introduced in \cite{Engels:1999wf} is unstable
precisely in this region. As a result, our determination of
$z_p$ is very precise in comparison to the previous estimates
from the interpolated form and the perturbative equation of state.

\section*{Acknowledgments}

The research of A.C.\ and T.M.\ is supported by FAPESP
(Project No.\ 00/05047-5).

\clearpage

\end{document}